\documentclass[aps,pre,twocolumn,groupedaddress]{revtex4}
\usepackage{amsfonts} \usepackage{amsmath}
\usepackage{graphicx}

\begin{document}

\title{Modulated spin waves and robust quasi-solitons in classical Heisenberg rings}



\author{Heinz-J\"urgen Schmidt}
\email[Electronic address: ]{hschmidt@uos.de}
\affiliation{Department of Physics, University of Osnabr\"uck, D-49069 Osnabr\"uck, Germany
}

\author{Christian Schr\"oder}
\affiliation{Department of Electrical Engineering and Computer Science,
University of Applied Sciences Bielefeld, D-33602 Bielefeld, Germany
\&
Ames Laboratory, Ames, Iowa 50011, USA
}

\author{Marshall Luban}
\affiliation{Ames Laboratory \& Department of Physics and Astronomy, Iowa State University, Ames, Iowa 50011, USA
}


\begin{abstract}
We investigate the dynamical behavior of finite rings of classical
spin vectors interacting via nearest-neighbor isotropic exchange in
an external magnetic field. Our approach is to utilize the solutions
of a continuum version of the discrete spin equations of motion (EOM)
which we derive by assuming continuous modulations of spin wave
solutions of the EOM for discrete spins. This continuous EOM
reduces to the Landau-Lifshitz equation in a particular limiting regime.
The usefulness of the continuum EOM is demonstrated by the fact that the
time-evolved numerical solutions of the discrete spin EOM closely
track the corresponding time-evolved solutions of the continuum equation.
Of special interest, our continuum EOM possesses soliton solutions,
and we find that these characteristics are also exhibited by the
corresponding solutions of the discrete EOM.
The robustness of solitons is demonstrated by considering cases where
initial states are truncated versions of soliton states and by
numerical simulations of the discrete EOM equations when the spins are
coupled to a heat bath at finite temperatures.
\end{abstract}


\maketitle


\section{Introduction\label{sec:I}}

With the remarkable progress
in recent years in synthesizing and
analyzing magnetic molecules of great diversity
\cite{GSV}-\cite{W2}
it is timely to focus on the dynamical behavior
of finite arrays of interacting spins.
In this paper we direct our attention to ring structures
consisting of $N$ equally-spaced (spacing $a$) classical spins (unit vectors)
that interact via ferromagnetic nearest-neighbor isotropic exchange and
are subject to an external magnetic field
$\mathbf{B} = B \mathbf{e}$.
The equations of motion (EOM) of the discrete spin vectors
can be written as
\begin{eqnarray}\label{I1}
\frac{d}{dt}\mathbf{s}_n(t) &=&
\mathbf{s}_n(t) \times
(\mathbf{s}_{n-1}(t)+\mathbf{s}_{n+1}(t)
+\mathbf{B}),\\ \nonumber
&&\quad n=0,\ldots,N-1,
\end{eqnarray}
where all variables are dimensionless
and the unit vectors $\mathbf{s}_n$ are
subject to the cyclic condition
$\mathbf{s}_{n+N}\equiv\mathbf{s}_{n}$.
It is well known that the solutions of Eq.~(\ref{I1})
exhibit a wide variety of dynamical characteristics.
Among these are:
Exact analytical solutions, called ``spin waves" \cite{mattis};
numerical solutions that exhibit chaotic behavior
\cite{schroeder};
as well as numerical solutions that are discretized
versions of the soliton solutions \cite{fogedby} of a continuum
version of Eq.~(\ref{I1}), specifically the
Landau-Lifshitz equation \cite{LL} in the case of $B\neq 0$.\\

The key result of this paper is that the EOM of
Eq.~(\ref{I1}) admit a wide class of numerical
solutions with soliton-like characteristics.
They are discretized versions of soliton solutions
of a new continuum EOM that includes the Landau-Lifshitz
equation as a special, limiting case.
We will refer to such discretized solutions as
``quasi-solitons", using this cautious terminology
since we do not know whether they correspond to
strict
soliton solutions of Eq.~(\ref{I1}).
A single quasi-soliton solution is a localized
disturbance that propagates in the ring while
for the most part maintaining its initial shape.
We also find that it is possible to achieve
two-quasi-soliton solutions, where a pair of
initially separated single quasi-solitons
collide multiple times and emerge from collisions
without appreciable modification. Moreover, we identify
solutions of Eq.~(\ref{I1}) that exhibit a transition from
laminar-like to turbulent-like behavior.\\

Our continuum EOM is obtained as follows.
We recall the general spin wave solution of Eq.~(\ref{I1})

\begin{equation}\label{I2}
\mathbf{s}_n(t) =
\left(
\begin{array}{l}
\sqrt{1-z^2}\cos(q n -B t -\omega t)\\
\sqrt{1-z^2}\sin(q n -B t -\omega t)\\
z
\end{array}
\right)
,
\end{equation}
where
\begin{equation}\label{I3a}
q=\frac{2\pi k}{N},\quad (k=0,\ldots,N-1)
\end{equation}
the angular frequency $\omega$ is given by
\begin{equation}\label{I3b}
\omega = 2(1-\cos q) z
\end{equation}
and $z$ is an arbitrary real number in the interval $-1 < z < 1$.
In Eq.~(\ref{I2}) we replace $z$ and $\omega t$ by two sets of
site-dependent functions $z_n(t)$ and $\varphi_n(t)$,
respectively, which are assumed to vary
slowly over a distance $a$.
As shown in Sec.~\ref{sec:D},
in a second order approximation scheme we obtain an EOM of the form

\begin{eqnarray}\nonumber
\frac{\partial\mathbf{S}}{\partial t} &=& \left[ B+(1-\cos q)\;(\mathbf{e}\cdot
(2\mathbf{S}+a^2 \frac{\partial^2\mathbf{S}}{\partial x^2})
\right] \mathbf{S}\times\mathbf{e}\\ \nonumber
&& -2a\sin q
\;(\mathbf{e}\cdot\mathbf{S})\;\frac{\partial\mathbf{S}}{\partial  x}\\ \label{I4}
&& +a^2\; \cos q \;\mathbf{S}\times
\frac{\partial^2\mathbf{S}}{\partial x^2} \;.
\end{eqnarray}
In the limit $q \rightarrow 0$, on rescaling the coordinate $x$,
this equation reduces to the Landau-Lifshitz equation,
\begin{equation}\label{I5}
\frac{\partial\,\mathbf{S}}{\partial\,t}
=
\mathbf{S}\times
\left(
 \frac{\partial^2\,\mathbf{S}}{\partial\,x^2}
+ \mathbf{B}
\right)
\;,
\end{equation}
and hence Eq.~(\ref{I4}) can be considered as the generalization of Eq.~(\ref{I5}).
As we show below Eq.~(\ref{I4}) possesses one-soliton solutions,
even if $B$ equals $0$, and probably also ${\cal N}$-soliton solutions.\\

The rationale for our strategy of developing
Eq.~(\ref{I4}) is as follows:
The replacement of the finite difference expression
$\mathbf{s}_{n+1} + \mathbf{s}_{n-1} - 2\mathbf{s}_n$ in
Eq.~(\ref{I1}) by the single term
$\frac{\partial^2\,\mathbf{S}}{\partial\,x^2}$, as in Eq.~(\ref{I5}),
is accurate as long as the
continuous function $\mathbf{S}(x,t)$ varies sufficiently
slowly over a distance $a$ so that the effects of
an infinite number of higher order spatial
derivatives can be neglected. In particular,
the solutions of the Landau-Lifshitz equation
can only describe small deviations from spin wave
solutions with $q$ approximately zero.
By contrast Eq.~(\ref{I4}) starts from the expressions of Eq.~(\ref{I2})
for any choice of $q$ allowed by Eq.~(\ref{I3a}).\\

A first order version of the continuum EOM of Eq.~(\ref{I4}),
as shown in Sec.~\ref{sec:FO}, can be solved completely.
It solutions do not describe solitons but deformations
of the initial profile similar to those of the
Hamilton-Jacobi equation for free particles.
After a finite time caustics are necessarily
formed indicating that the
first order approximation breaks down.
In our numerical studies of  Eq.~(\ref{I1}) we do find
instances where smooth or "laminar" solutions
give way to a chaotic or "turbulent" regime,
but this does not occur in general.\\

Solitons appear in the second order approximation of the
EOM of Eq.~(\ref{I4}).
Using a discretized version of the initial
form of a solution of Eq.~(\ref{I4}),
we find that, under certain conditions,
the time-evolved numerical solution of the discrete spin
EOM closely tracks the time-evolved solution.
We call these numerical solutions ``quasi-solitons".
However, we have found that, in the case of $B=0$,
the existence of quasi-solitons
is limited to a small ``window" of $q$-values, say
$\frac{6\pi}{N}<|q|<\frac{20\pi}{N}$ for $N=100$.
Were we not guided
by the solutions of the continuum equation,
the chance of successfully generating a
quasi-soliton solution to the discrete EOM
is much like the proverbial
``finding a needle in the haystack".
In support of this, we find that for virtually
all other choices of an initially localized
spin configuration the spin pattern spreads
with time and does not reassemble, let alone
maintain its localized shape.
This also applies to the first order approximation
of the EOM, where
we have also found numerical solutions that
closely follow the initial stages of the
evolution towards a caustic.\\

It is remarkable that the dynamical behavior
of a relatively small number of interacting
spins is so multi-faceted and unexpectedly rich,
and that one is dependent on a study of the
continuum version of the EOM to provide the
key for unraveling and elucidating the
dynamical properties of the discrete EOM.
We have not attempted to analyze the
third order version of the continuum EOM
due to the severe mathematical complications
that arise. Other interesting dynamical behavior,
as yet unknown to us, might still be found
from a deeper mathematical analysis of the
complete continuum EOM.\\

Another important aspect is the question of
how robust quasi-solitons are. One possible
perturbation of solitons is the coupling of the spins to a heat bath.
This kind of thermal perturbation is incorporated in
our numerical treatment of the discrete
EOM in Sec.~\ref{sec:HB}. Likewise, we explore rings
with relatively small values of $N$. We find
that quasi-soliton solutions are relatively
robust, surviving as long-lived spatially
localized patterns as the temperature is
increased and/or the initial profile is
truncated or as $N$ is decreased to as small as $11$.\\

The kind of quasi-solitons investigated in the present paper
represent only a certain fraction of solitary phenomena
in spin systems known from the literature.
Another class of solitons exists for
classical chains with a particular
form of anisotropy that leads to the
Sine-Gordon equation \cite{mikeska}.
Other soliton solutions for continuous EOM
describing similar systems have been given
in \cite{YCH}-\cite{GGI}.
Solitary solutions of a quantum spin chain have been
investigated, for example, in \cite{SS}.
A very large literature
(see \cite{mikeska} for a comprehensive review)
exists which has shown that these EOM
are of direct relevance to a wide class
of one-dimensional magnetic materials and
provide a variety of theoretical models in quantum field theory.\\

The layout of this paper is as follows.
In Sec.~\ref{sec:D}, starting from the discrete
EOM for the interacting spin system, Eq.~(\ref{I1}),
we review the well-known exact spin-wave
solutions, and then proceed to derive a
continuum version of the EOM describing modulated
spin waves. Sec.~\ref{sec:FO} is devoted
to deriving the full set of solutions of the
first order version of the continuum EOM,
and in particular to showing that any
solution necessarily develops a caustic
in the course of time. We include examples
of numerical solutions of Eq.~(\ref{I1})
showing such behavior.
In Sec.~\ref{sec:SO} we give a
detailed analysis of the soliton solutions of the
second order version of the continuum EOM.
We obtain a formula for the dependence of the
soliton amplitude $A$ on the wave number $q$, the
soliton velocity $u$, and the magnetic field $B$.
In the limit $q\rightarrow 0$ this formula reduces to
the well-known result for the Landau-Lifshitz
equation. In Sec.~\ref{sec:NA} we present results of
our numerical analysis of the discrete EOM of Eq.~(\ref{I1}).
These include one-quasi-solitons and multiple-quasi-solitons
starting from initial profiles
obtained from soliton solutions of the
second order continuum EOM.
Sec.~\ref{sec:HB} is devoted to the effects of starting
from initial spin configurations that are
truncated versions of soliton states, as
well as including the coupling of the
discrete spins to a heat bath.
Finally, in Sec.~\ref{sec:SU} we summarize our
results and discuss open questions.\\


\section{Modulated spin waves\label{sec:D}}


We consider classical spin rings with $N$ spins interacting via
ferromagnetic nearest-neighbor coupling.
The equations of motion are given by Eq.~(\ref{I1})
where $\mathbf{B}=B \mathbf{e}$ is the dimensionless magnetic field and
$\mathbf{e}$ is chosen as a unit vector in the direction of the $3$-axis.
Mathematically, the case of anti-ferromagnetic coupling is included
since it can be achieved by the transformation $\mathbf{s}_n\mapsto -\mathbf{s}_n$.
However, solitons which are excitations of a background of fully aligned spins
will be thermodynamically stable only in the ferromagnetic case. Hence we will stick to
the ferromagnetic case for physical reasons.
For $N>4$ the general solution of Eq.~(\ref{I1}) cannot be given in closed form.
However, as mentioned above, there exist spin wave solutions of the form of Eq.~(\ref{I2}).
These solutions are parameterized by a variable $z$, the common $3$-component
of all spin vectors, and by the wave number $q$ which assumes the discrete
values of Eq.~(\ref{I3a}).\\

In this article we will consider modulations of Eq.~(\ref{I2}) where $z$ and $\omega t$
are replaced by two sets of functions $z_n(t)$ and $\varphi_n(t)$, respectively, which are
assumed to vary slowly over a distance $a$. That is,
$\mathbf{s}_n(t)$ is given by

\begin{eqnarray}\label{D1}
\mathbf{s}_n(t) &=&
\left(
\begin{array}{l}
\sqrt{1-z_n(t)^2}\cos(q n -B t-\varphi_n(t))\\
\sqrt{1-z_n(t)^2}\sin(q n -B t-\varphi_n(t))\\
z_n(t)
\end{array}
\right)
\end{eqnarray}
It is convenient to remove the term $qn$  by the
transformation
\begin{equation} \label{D2}
\widetilde{\mathbf{s}}_n(t)\equiv R_3(-\,q\,n)\;\mathbf{s}_n(t)
\;,
\end{equation}
where $R_3(\alpha)$ denotes the matrix of a rotation about the $3$-axis with an angle $\alpha$.
One could also remove the magnetic field by a suitable uniform rotation but we prefer to retain $B$
in what follows in order to make the transition to the Landau-Lifshitz equation more transparent.
\\

In the continuum approximation we introduce smooth functions
$z(x,t), \varphi(x,t)$ and $\mathbf{S}(x,t)$ such that
$z(na,t)=z_n(t)$ and $\varphi(na,t)=\varphi_n(t)$, and
\begin{eqnarray}\nonumber
\mathbf{S}(x,t) &=&
\widetilde{\mathbf{s}}_{x/a}(t)
=
\left(
\begin{array}{l}
\sqrt{1-z(x,t)^2}\cos(B t+\varphi(x,t))\\
-\sqrt{1-z(x,t)^2}\sin(B t+\varphi(x,t))\\
z(x,t)
\end{array}
\right)\\ \label{D3}
&&
\end{eqnarray}
In the continuum approximation we allow for the parameter $q$ to take on any values in the
interval $(-\pi,\pi)$, whereas the Landau-Lifshitz equation is obtained in the limit
$q\rightarrow 0$.\\

Using the addition theorems for $\cos$ and $\sin$ we obtain
\begin{equation}\label{D4}
\mathbf{s}_{n\pm 1} =
\left(
\begin{array}{l}
\sqrt{1-z_{n\pm 1}^2}\left[\cos\alpha_\pm \cos q \mp \sin\alpha_\pm \sin q\right]\\
\sqrt{1-z_{n\pm 1}^2}\left[\sin\alpha_\pm\cos q \pm \cos\alpha_\pm\sin q\right]\\
z_{n\pm 1}(t)
\end{array}
\right)\\
\end{equation}
where
\begin{equation}\label{D4a}
\alpha_\pm \equiv qn-Bt-\varphi_{n\pm 1}(t)
\,.
\end{equation}

Applying the rotation $R_3(-\,q\,n)$ to $\mathbf{s}_{n+1}+\mathbf{s}_{n-1}$ hence yields
\begin{eqnarray}\nonumber
R_3(-\,q\,n)(\mathbf{s}_{n+1}+\mathbf{s}_{n-1}) &=&\\ \nonumber
(\widetilde{\mathbf{s}}_{n+1}+\widetilde{s}_{n-1})\cos q &&\\ \nonumber
+\mathbf{e}\times(\widetilde{\mathbf{s}}_{n+1}-\widetilde{s}_{n-1})\sin q&&\\ \label{D5}
+(z_{n+1}+z_{n-1})(1-\cos q)\mathbf{e}.&&
\end{eqnarray}

Finally, we use the representation of Eq.~(\ref{D3}) and
apply the rotation $R_3(-\,q\,n)$ to the EOM of Eq.~(\ref{I1})
thus obtaining the continuum approximation of the form
\begin{equation}\label{D6}
\frac{\partial}{\partial t}\mathbf{S}(x,t)=\mathbf{S}(x,t)\times\mathbf{H}(x,t),
\end{equation}
where
\begin{eqnarray}\nonumber
\mathbf{H}(x,t)=
\left[
(\mathbf{S}(x+a,t)+\mathbf{S}(x-a,t))
\right]\cos q
&&\\ \nonumber
+\mathbf{e}\times\left[
\mathbf{S}(x+a,t)-\mathbf{S}(x-a,t)
\right]\sin q
&&\\ \label{D7}
+\left[
B +
\left(z(x+a,t)+z(x-a,t)\right)
(1-\cos q)\right]\;\mathbf{e}
.&&
\end{eqnarray}

In the following we consider truncated versions of Eqs.~(\ref{D6}) and (\ref{D7}).
These are obtained by first expanding $\mathbf{S}(x\pm a,t)$ in powers of
$a$. We obtain an infinite order partial differential equation (PDE) of the
form of Eq.~(\ref{D6}) with
\begin{eqnarray}\nonumber
&&\mathbf{H} =
2\left(
\mathbf{S}+\frac{a^2}{2}\mathbf{S}''+\frac{a^4}{4!}\mathbf{S}^{(4)}+\ldots
\right)\cos q \\ \nonumber
&& +2 \mathbf{e}\times\left(
a\mathbf{S}'+\frac{a^3}{3!}\mathbf{S}^{(3)}+\ldots
\right)\sin q \\ \nonumber
&& +\left[
B + 2\left(
z+\frac{a^2}{2}z''+\frac{a^4}{4!}z^{(4)}+\ldots
\right)(1-\cos q)\right]
\;\mathbf{e}
,\\ \label{D8}
&&
\end{eqnarray}
where the arguments $(x,t)$ have been suppressed, and we abbreviate the
spatial derivatives of $\mathbf{S}$ and $z$ by a prime and the time
derivatives by a dot in what follows.
In the $n$th order approximation we keep terms containing powers of $a$ through $a^n$.
In this article we consider zeroth, first, and second order approximations only.
The $0$th order PDE is given by
\begin{equation}
\dot{\mathbf{S}}=(B+2z(1-\cos q))\mathbf{S}\times\mathbf{e}
\,.
\end{equation}
This equation is solved by the spin wave solution, Eq.~(\ref{I2}),
but it also has more general solutions.

\section{First order equations\label{sec:FO}}

In the first order approximation Eqs.~(\ref{D6}) and (\ref{D8}) reduce to
\begin{equation} \label{FO1}
\dot{\mathbf{S}}=\left[
B+2(1-\cos q)\;(\mathbf{e}\cdot\mathbf{S})\right]
\;\mathbf{S}\times\mathbf{e}
-2a\sin q \;(\mathbf{e}\cdot\mathbf{S})\;\mathbf{S}'
\;.
\end{equation}
Using the representation of Eq.~(\ref{D3}) we can rewrite this equation as
\begin{eqnarray}\label{FO2a}
\dot{\varphi}&=&B+2 z(x,t)((1-\cos q) - a \sin q \;\varphi')\\ \label{FO2b}
\dot{z}&=&-2 a \sin q \;z\; z'
\;.
\end{eqnarray}
First we consider Eq.~(\ref{FO2b}), which is an autonomous equation for $z(x,t)$.
Its solution is well-known, see e.~g.~\cite{whitham}.
Equation (\ref{FO2b}) essentially describes the velocity field of a system of free particles
with a given velocity distribution for $t=0$. To make this more transparent, consider the
transformation
\begin{figure}
  \includegraphics[width=\columnwidth]{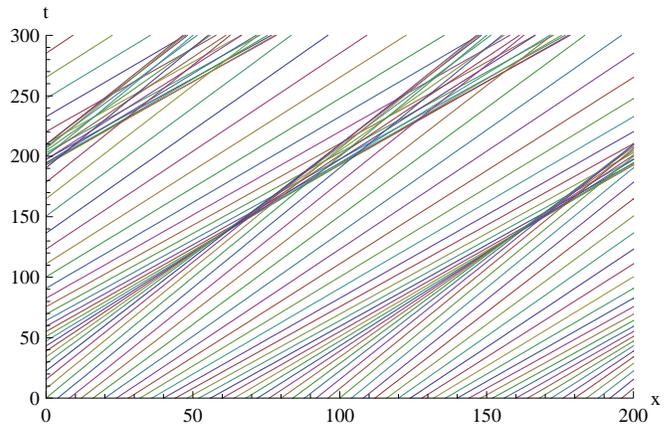}
\caption{\label{fig01}Caustics in a line field with initial slope
$v(x,0)=\frac{1}{2}+\frac{1}{4}\sin^2(\frac{\pi x}{100})$.
}
\end{figure}
\begin{equation}\label{FO3}
v(x,t)\equiv z(x,\frac{t}{2a\sin q}),
\end{equation}
which transforms Eq.~(\ref{FO2b}) into
\begin{equation}\label{FO4}
\dot{v}+ v v'=0
\;.
\end{equation}
Hence $v$ is constant along the lines with slope
$\frac{dx}{dt}=v$ whence the above kinematical interpretation follows.
Let $\xi= x - u t$ be the $x$-coordinate of the intersection of the axis $t=0$
and the line through the point $(x,t)$ with slope $u$.
Hence $u=v(\xi,0)$. This yields a parametric representation
of the graph of the general solution $x\mapsto v(x,t)$ of
Eq.~(\ref{FO4}) for fixed $t$, namely
\begin{eqnarray}\label{FO5a}
x&=&\xi + v(\xi,0)t\\ \label{FO5b}
u&=&v(\xi,0)
\;,
\end{eqnarray}
where $v(\xi,0)$ is an arbitrary initial condition for the considered PDE. An example of the resulting line
field is given in Fig.~\ref{fig01}. We see that there are domains, where the lines in the line field intersect,
bounded by curves called ``caustics". These domains correspond to points where the
parametric representation of Eqs.~(\ref{FO5a}), (\ref{FO5b}) defines a multi-valued function.
The earliest time when caustics appear is given by
\begin{equation}\label{FO6}
t_c=-\frac{1}{\frac{\partial v(\xi_0,0)}{\partial \xi}},
\end{equation}
where $\xi_0$ is the point with the largest negative derivative of $v(\xi,0)$. Initial profiles $v(\xi,0)$
with only positive slope do not have caustics for $t>0$ but cannot be realized on a spin ring with
periodic boundary conditions.\\

\begin{figure}
  \includegraphics[width=\columnwidth]{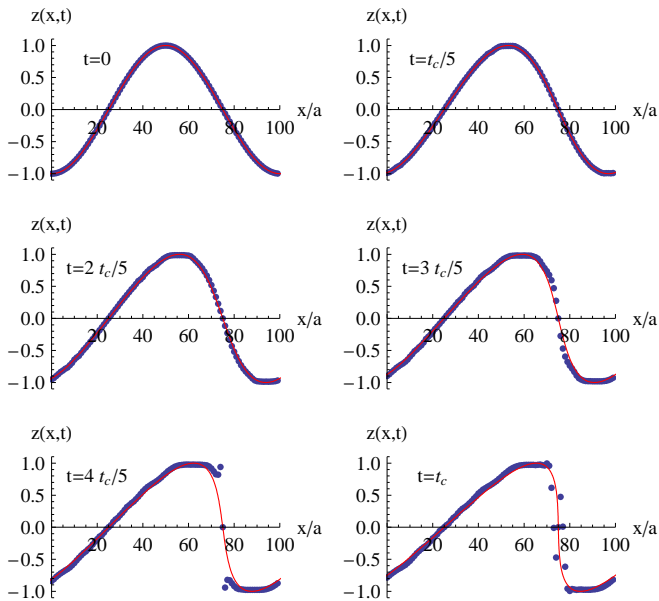}
  \caption{\label{fig02}Numerical solutions $z_n(t)$ of the discrete EOM of Eq.~(\ref{I1})
for a spin ring with $N=100$ (blue dots) and solution of the $1$st order continuum
approximation $z(x,t)$ (solid red curve), where $t=n t_c/5,\, n=0,1,\ldots,5$ and
$t_c$ is the caustic time defined in Eq.~(\ref{FO6}).
For $t>t_c$ the solid curve bends over and defines a $3$-valued function.}
\end{figure}

The same analysis applies to the
original equation Eq.~(\ref{FO2b}), except that the caustic time $t_c$ in
Eq.~(\ref{FO6}) has to be divided by $2 a \sin q$.
Hence the first order PDE predicts that after some time $t_c$ the profile of the $3$-components of the spin ring
develops an infinite slope. That is for later times the $1$st order analysis is no longer valid and the
equations of motion must be studied by using higher order approximations. Remarkably, there exist examples
where the numerical solution of Eq.~(\ref{I1}) is well described by its $1$st order approximation for times $t$ with
$0\le t < t_c$. Close to the time $t=t_c$ the smooth ``laminar" time evolution breaks down and a different regime begins to
penetrate the spin ring. We will call this regime ``turbulent" for sake of simplicity, but we do not know
whether it is genuinely chaotic or still regular on a shorter length scale.
From our numerical results the latter cannot be excluded.
Figure \ref{fig02} shows the numerical solution $z_n(t)$ of Eq.~(\ref{I1})
for an $N=100$ spin ring with $q=16\pi/100$, together with the $1$st order PDE solution $z(x,t)$ at the time $t=t_c$.
The two sets of results are in close agreement, although at time $t=t_c$ the continuum approximation is
no longer applicable.

We remark that equation (\ref{FO2a}) can also be solved in closed form. Its general
solution utilizes the solution $z(x,t)$ of Eq.~(\ref{FO2b})
already obtained and the initial angle distribution $\varphi_0(x)=\varphi(x,0)$ and it is given by
\begin{equation}\label{FO7}
\varphi(x,t)=\varphi_0(x-2a\sin q \; z(x,t) t)+z(x,t)2(1-\cos q)t+Bt
\;.
\end{equation}
Hence the effects of the caustics will also be manifested in the behavior
of the azimuthal angles of the spin vectors.

\section{The second order equations\label{sec:SO}}

In the preceding section we have seen that the $1$st order approximation predicts its own break-down
since it will necessarily lead to a caustic or ``turbulent" behavior. Hence the question arises whether higher
order approximations give rise to the same or different effects. The second order equation
is given by
\begin{eqnarray}\nonumber
\dot{\mathbf{S}}&=&
\left[
B+2(1-\cos q)(\mathbf{e}\cdot\mathbf{S})\right]
\mathbf{S}\times\mathbf{e}\\ \nonumber
&-& 2 a \sin q (\mathbf{e}\cdot\mathbf{S})\mathbf{S}'\\ \label{SO1}
&+&  a^2 \mathbf{S}\times
\left[\cos q\,\mathbf{S}''+(1-\cos q)(\mathbf{e}\cdot\mathbf{S})\mathbf{e}
\right].
\end{eqnarray}

Substituting Eq.~(\ref{D3}) leads to
\begin{eqnarray}\nonumber
\dot{\varphi}&=&B+2 z\;\left(1-\cos q - a \sin q \;\varphi'\right)\\ \nonumber
&+&  a^2 \left[
z''\,+z\cos q\;\left(
\varphi'^2+\frac{z z''}{1-z^2}+\frac{z'^2}{(1-z^2)^2}
\right)\right]\\ \label{SO2a}
&&\\
\nonumber
\dot{z}&=&-2 a \;\sin q \;z\; z'\\  \label{SO2b}
&&+ a^2 \cos q(2 z\; z'\;\varphi' -(1-z^2) \;\varphi'').
\end{eqnarray}

\begin{figure}
\includegraphics[width=\columnwidth]{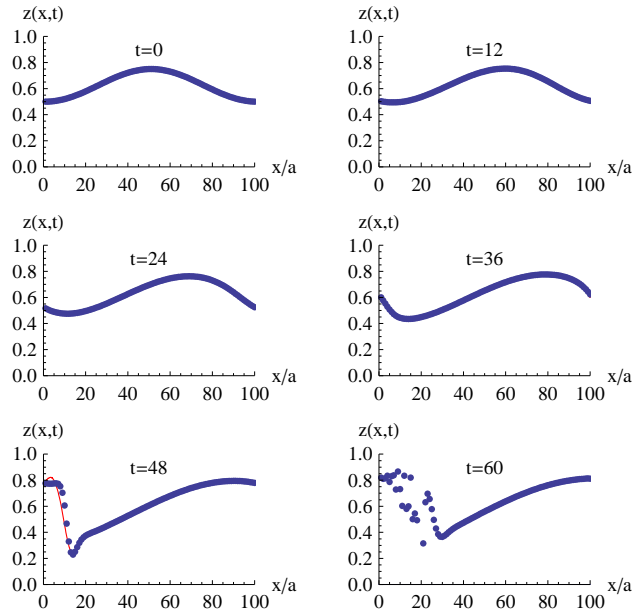}
\caption{\label{fig03}Comparison of a numerical
solution for $z_n(t)$ at fixed times $t=0,12,\ldots,60$
of the equations of motion for an $N=100$
spin ring (blue dots) with the numerical solution of its
$2$nd order continuum approximation Eqs.~(\ref{SO2a}), (\ref{SO2b})
(solid red curve) at the times $t=0,12,\ldots,48$.
The turbulent regime begins to spread at approximately $t=48$. By $t=60$ the numerical solution breaks down
and hence cannot be drawn.
}
\end{figure}

In solving Eq.~(\ref{SO1}) numerically and analytically we have found instances where the onset of
turbulence is accelerated as well as where it is entirely suppressed. We do not, however, have a
clear understanding about the underlying causes for either behavior. As an example of the first case,
acceleration of turbulence, in Fig.~\ref{fig03} we compare results of the numerical solution
of the EOM of Eq.~(\ref{I1}) with the numerical solution of Eqs.~(\ref{SO2a}) and (\ref{SO2b}).
The time of the onset of the turbulence is approximately 2.75 times shorter than that
predicted by the $1$st order approximation.

The other case, preservation of laminar behavior, due to $2$nd order effects, can be most impressively demonstrated
by the existence of solitons, presented in the next subsection.\\

\subsection{Exact single soliton solutions\label{sec:E}}

The mathematical derivation of soliton solutions is to a large extent analogous to the
Landau-Lifshitz case, see \cite{mattis}, \cite{fogedby}.
To simplify the derivation we set $a=1$ in the following.\\
We seek solutions of Eqs.~(\ref{SO2a}) and (\ref{SO2b}) of the form
\begin{eqnarray}\label{E1a}
\varphi(x,t)&=&\Phi(x-u\;t)\\ \label{E1b}
z(x,t)&=&Z(x-u\;t)
\;.
\end{eqnarray}
The resulting equation for $Z(x)$
can be integrated once yielding
\begin{equation}\label{E2}
\Phi'=\sec q\;\frac{u(Z-Z_0)- Z^2\;\sin q}{1-Z^2}
\;,
\end{equation}
where $Z_0$ is some integration constant. Substituting this into the equation resulting from Eq.~(\ref{SO2a})
yields a second order equation for $Z(x)$. Multiplying with $Z'(x)$ and integrating
with respect to $x$ yields
an expression for $Z'(x)^2$ containing an integration constant $c$.
This expression is analogous to a one-dimensional potential if $Z'(x)^2$ is viewed as
the kinetic energy and the total energy is set equal to $0$, namely
\begin{eqnarray}\nonumber
Z'^2&=&-V(Z)=\frac{1}{2\, (1-Z^2(1-\cos q))}\\ \nonumber
&&\left\{
\sec q
\left[
-1-3Z^2-2u^2(2ZZ_0-1-Z_0^2)\right.\right.\\ \nonumber
&& -(1+Z^2)\cos 2q +8Z^2(1-Z^2)\sin^4\frac{q}{2}\left.\right]\\ \nonumber
&&+4\left[
c(Z^2-1)+Z(Z+B(1-Z^2))
 \right.\\ \label{E3}
&&+u  (Z_0-Z^3)\tan q \left.\right]\left.\right\}
\;.
\end{eqnarray}
$V(z)$ will be called the pseudo-potential.
Since $-V(Z)$ is  biquadratic in $Z$ in the numerator and quadratic in the denominator the integral
\begin{equation}\label{E4}
x-x_0 = \int\frac{dZ}{\sqrt{-V(Z)}}
\end{equation}
can be expressed in terms of elliptic functions of the first and third kind, see \cite{AS}. The inverse
function of Eq.~(\ref{E4}), together
with the integral of Eq.~(\ref{E2}), gives the inverse function of the soliton solution.
However, the analytic expression is very complicated and it is far simpler to calculate the integral
of Eq.~(\ref{E4}) numerically. The poles of $V(Z)$ lie at $Z_p= \frac{\pm 1}{\sqrt{1-\cos q}}$
and hence are outside the physical domain $Z\in [-1,1]$ for $0<|q|<\frac{\pi}{2}$.
Hence we will assume
\begin{equation}\label{E5}
0<|q|<\frac{\pi}{2}
\end{equation}
in the following.
\\

The remaining task is to determine values of the parameters $q,B,u,Z_0,c$ for which solitary solutions
of Eqs.~(\ref{SO2a}) and (\ref{SO2b}) exist.
The inverse function of (\ref{E4}) has the form of a soliton profile if
$V(Z)$ has a double root at $Z=Z_1$ and a simple root at $Z=Z_2$ such that $-1\le Z_1<Z_2\le 1$
and $V(Z)$ is negative in the interval $(Z_1,Z_2)$. For sake of simplicity we only consider the case
$Z_1=-1$.\\

\begin{figure}

 \includegraphics[width=\columnwidth]{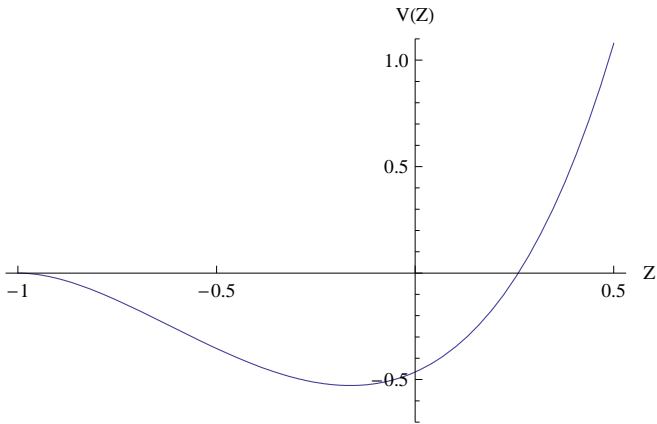}
\caption{\label{fig04}The pseudo-potential $V(Z)$ according to Eq.~(\ref{E3})
corresponding to the parameters $B=0,\,q=-\pi/4$ and $u=1$.
}
\end{figure}

The condition that $V(Z)$ has a double root at $Z=Z_1=-1$ gives us two parameters as a function of the
remaining two:
\begin{eqnarray}\label{E6a}
c&=& 1-B -\frac{(2+Z_0)^2\cos q+(Z_0^2-2)\sec q}{4\cos q\;(1+Z_0)^2}
\\ \label{E6b}
u&=&-\frac{\sin q}{1+Z_0}
\;.
\end{eqnarray}
The soliton solutions hence depend on three parameters which we choose to be $B,\,q$ and $u$.
A typical form of $V(Z)$ is shown in Fig.~\ref{fig04}. For the numerical studies in Sec.~\ref{sec:NA} and \ref{sec:HB}
we mostly chose negative wave-numbers $q$ in order to get positive velocities according to Eq.~(\ref{E6b}).

The remaining roots $Z_{2,3}$ are
\begin{eqnarray} \nonumber
Z_{2,3}&=&
\frac{1}{8}\csc^4\frac{q}{2}\;
\left[
3+\cos 2q+2u\sin q
\right.\\ \label{E7}
&&-2 \cos q\;(2-B\mp f(B,q,u))\left.\right]\\ \nonumber
\mbox{where}&&\\ \label{E8}
f(B,q,u)&\equiv&\sqrt{B^2-2u^2+2u\sec q\,(u+B\sin q)}
.
\end{eqnarray}

\begin{figure}
 \includegraphics[width=\columnwidth]{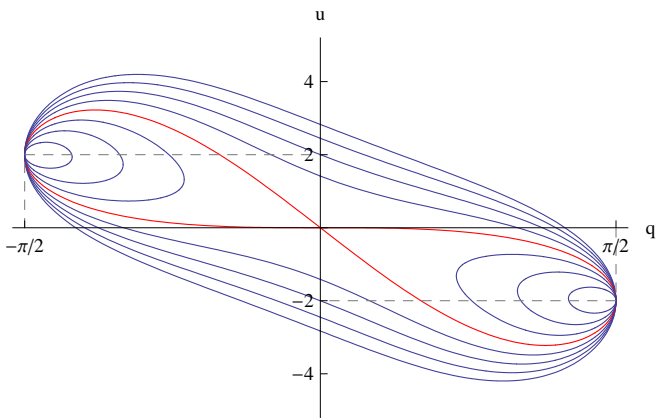}
\caption{\label{fig05}
The soliton velocities $u$ bounded by a function of $q$ and $B$ according to Eq.~(\ref{E9}).
Curves shown correspond to $B=-2,\,-1.5,\ldots,\,1.5$ The $\infty$-shaped red curve corresponds
to $B=0$. For $B\rightarrow 2$ the curves shrink to two points at $q=\pm \pi/2,\; u=\mp 2$.
}
\end{figure}

The condition that $V(Z)$ is negative in the interval $Z\in(-1,Z_2)$
or, equivalently, $V''(-1)<0$, which is necessary for the existence of solitons,
leads to the inequalities
\begin{equation}\nonumber
-2\sin q-2\cos q\sqrt{(2-B)\sec q-2}
\end{equation}
\begin{equation}\label{E9}
<u<
\end{equation}
\begin{equation}\nonumber
-2\sin q+2\cos q\sqrt{(2-B)\sec q-2}
\end{equation}
This defines a certain physical domain in the $(B,q,u)$-plane, see Fig.~\ref{fig05}, where solitons exist.
It is bounded by $B\le 2$. The inequalities further imply that only the root $Z_2$ of $V(Z)$ with the $-$-sign
in Eq.~(\ref{E7}) lies inside the physical domain $(-1,1)$.

\begin{figure}
 \includegraphics[width=\columnwidth]{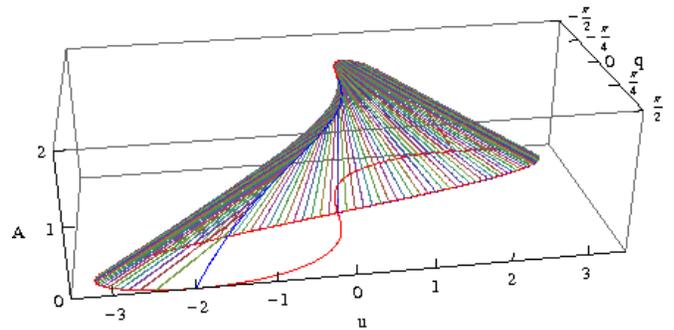}
\caption{\label{fig06}
Soliton amplitude $A$ as a function of wave number $q$ and velocity $U$ for $B=0$ according to Eq.~(\ref{E12}).
}
\end{figure}

Eq.~(\ref{E7}) yields an explicit relation between the amplitude $A\equiv Z_2-Z_1=1+Z_2$
and the velocity $u$ of the soliton of the form
\begin{equation}\label{E10}
A=2+\frac{u\sin q +(B+f(B,q,u))\cos q}{4 \sin^4\frac{q}{2}}
\;,
\end{equation}
with $f(B,q,u)$ defined by Eq.~(\ref{E8}). The two limiting forms are
\begin{equation}\label{E11}
A=2+\frac{u^2}{2B}\mbox{ for }q\rightarrow 0 \mbox{ and } B<0
\;,
\end{equation}
to be compared to Eq.~(4.7) in \cite{fogedby}, and
\begin{eqnarray}\nonumber
A&=&2+\frac{1}{4}\csc\frac{q}{2}\left(
u\sin q-|u|\sqrt{2(\sec q\,-1)}\cos q
\right)\\ \label{E12}
&& \mbox{for }B\rightarrow 0
\,.
\end{eqnarray}
The physical part of the surface in $(q,u,A)$-space spanned by the lines
defined by Eq.~(\ref{E12}) with $q=$ const.~is shown in Fig.~\ref{fig06}.
From these equations it follows that the amplitude $A$ diverges for
$B,q\rightarrow 0$ and hence no solitons exist in this case in accordance with
\cite{fogedby}.

\section{Numerical soliton solutions\label{sec:NA}}

\begin{figure}
 \includegraphics[width=\columnwidth]{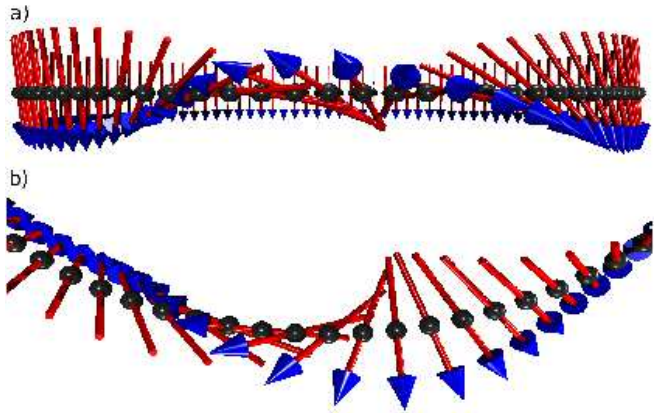}
\caption{\label{fig07}
Initial profile of a discrete soliton for a spin ring with $N=100$ calculated
from Eqs.~(\ref{E2}) - (\ref{E4}) for parameters $q=-8\pi/100$ and $u=\sin |q|$.
Panel a) shows the in-plane view while panel b) shows the structure of the soliton
as viewed from above. The initial profile $z_n(t)$ can be traced by eye following the
blue arrow tips. The time evolution of this soliton is shown in Fig.~\ref{fig08}.
}
\end{figure}

\begin{figure}
\includegraphics[width=\columnwidth]{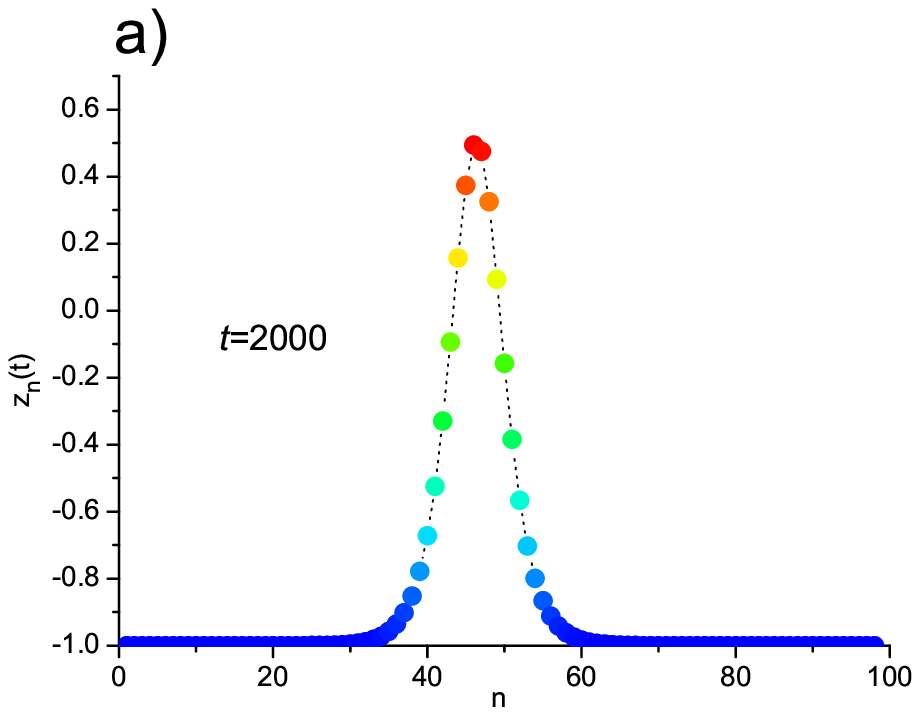}
\vspace{-5mm}
\includegraphics[width=\columnwidth]{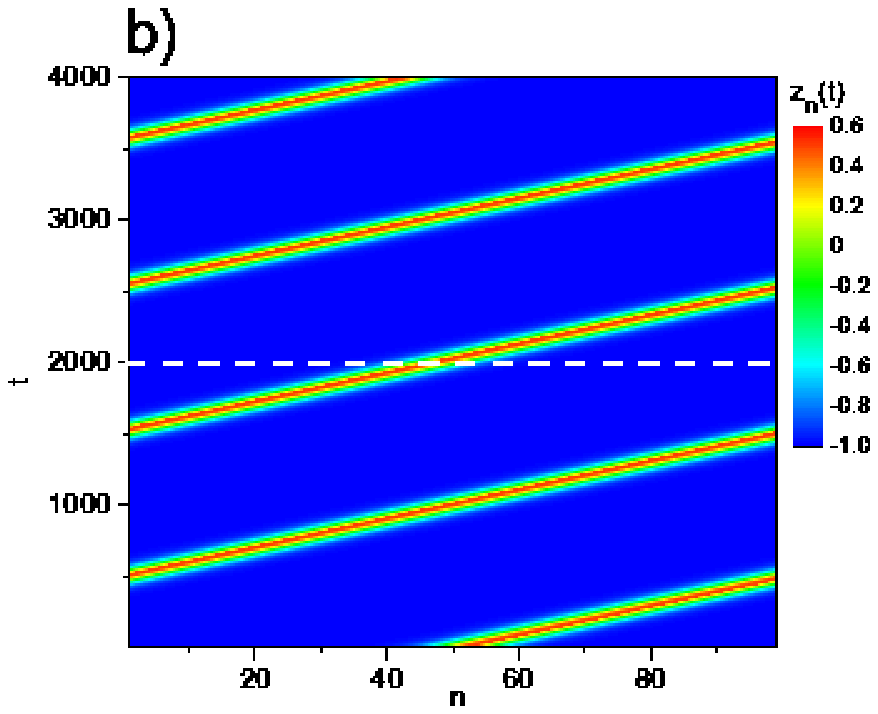}
\caption{\label{fig08}
Time evolution of the discrete soliton with initial profile shown in Fig.~\ref{fig07}.
Panel a) shows a snapshot of the values of $z_n(t)$ for $t=2000$. The color coding
corresponds to the contour plot given in panel b). The data given in panel a) correspond
to the dashed white line shown in panel b).
}
\end{figure}

In this section we summarize some of our results
for numerical solutions of the EOM, Eq.~(\ref{I1}), in the
case $B=0$, for finite spin rings. The initial spin
configurations are selected to be soliton profiles as
calculated from the $2$nd order continuum PDE using
Eqs.~(\ref{E2}) - (\ref{E4}). Fig.~\ref{fig07} shows the initial profile
of a discrete soliton for a spin ring with $N=100$ and
parameters $q=-8\pi/100$ and $u=-\sin q$. The soliton appears
as a localized deviation from an otherwise ferromagnetic
alignment of spins. Using this initial profile we solve
the EOM, Eq.~(\ref{I1}), numerically to determine the time
evolution of the spin vectors. The results are shown in
Fig.~\ref{fig08}. In panel a) we provide a snapshot of $z_n(t)$ for
a particular time $t=2000$. A comprehensive picture for
all calculated times is given in the color contour plot
of panel b), where the color gives the value of $z_n(t)$
according to the coding defined in the legend. The strict
integrity of the solution as a function of time is striking.\\
Throughout the remainder of this paper we use color
contour plots since they provide a very effective tool for
visualizing the overall time evolution of the spin profile
as measured by $z_n(t)$. In particular, these plots allow one to
detect even small changes in the profile, its speed, etc.\\

\begin{figure}
 \includegraphics[width=\columnwidth]{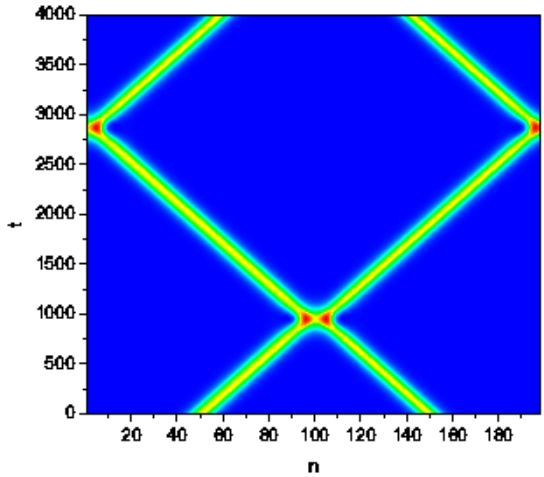}
\caption{\label{fig09}
Collision of two solitons of an $N=200$ spin ring. The initial profile
has been calculated from Eqs.~(\ref{E2}) - (\ref{E4}) using the parameters
$q=\pm 16\pi/200$ and $u=-\sin q$. The numerical solutions are displayed
for $t=0,\ldots,\,4000$, including the times $t_1\approx 1000$ and $t_2\approx 2800$
where collision between the two solitons take place.
}
\end{figure}

\begin{figure}
 \includegraphics[width=\columnwidth]{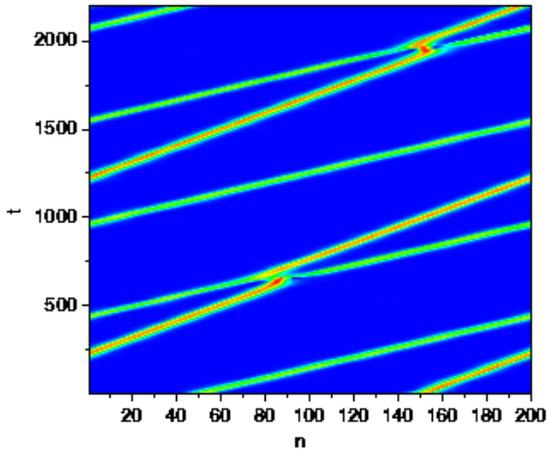}
\caption{\label{fig10}
Numerical two-soliton solution $z_n(t)$ for an $N=200$ spin ring
and initial data calculated from Eqs.~(\ref{E2}) - (\ref{E4}) using the parameters
$q=-16\pi/200$ and $u=0.625 \sin |q|$ resp.~$u=1.25 \sin |q|$. The solution is displayed
for $t=0,\ldots,\,2200$. In this case two solitons are created which move in the same direction
while one soliton is twice faster than the other eventually penetrating the slower one.
}
\end{figure}

Using the initial profile for a single soliton solution
it is easy to prepare the case of two identical solitons
moving towards each other. We copy the single soliton profile
of the $N=100$ spin ring onto one half of an $N=200$ spin ring
and its reversed version onto the second half of the spin ring.
As is seen in Fig.~\ref{fig09}, the numerical solution
of Eq.~(\ref{I1}) describes two solitons that collide and then move apart
while retaining their initial shapes and velocities. This behavior
is well-known for exact ${\cal N}$-soliton solutions of other
nonlinear wave equations, see, e.~g.~\cite{AS}. We have observed
repeated collisions of the soliton pairs without noticeable deformation. By
using a similar procedure we have created the situation
where a faster soliton overtakes a slower one, and the
results are shown in Fig.~\ref{fig10}. One can see that the
penetration of the slower soliton by the faster one does not
alter the shape or velocity of either soliton subsequent to their
collision.

\section{Effects of truncation and finite temperature \label{sec:HB}}

\begin{figure}
 \includegraphics[width=\columnwidth]{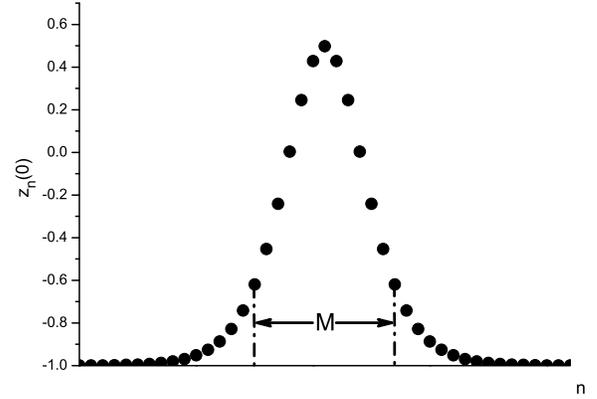}
\caption{\label{fig11}
Non-truncated initial profile (black dots) and truncated initial profile
which is obtained by taking out $M$ spins centered about the site
having the largest value of $z_n(0)$ and setting the value of $z_n(0)$
equal to $-1$ for the remaining spins.
}
\end{figure}

The purpose of this Section is to explore the extent
to which quasi-solitons are robust to various perturbations.
We begin by investigating the time evolution of
symmetrically truncated versions of initial states
that exhibit soliton behavior. This is illustrated
schematically in Fig.~\ref{fig11}, where the black dots
describe a hypothetical non-truncated initial profile.
In the truncated version, for all but $M$ spins
the value of $z_n(0)$ is set equal to $-1$, whereas we retain the original
values of $z_n(0)$ for the $M$ spins. The latter are
chosen to be centered about the site having the largest value of $z_n(0)$.\\

\begin{figure}
 \includegraphics[width=\columnwidth]{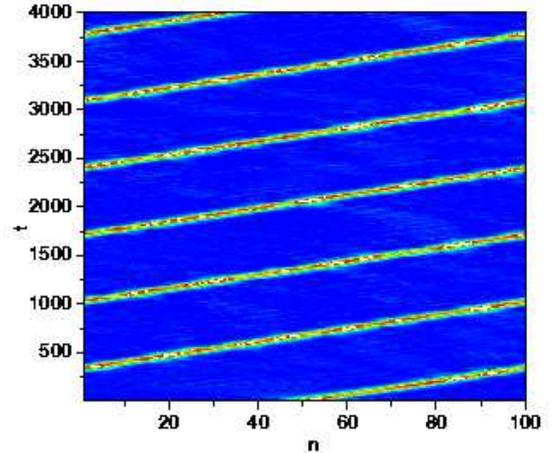}
\caption{\label{fig12}
Time evolution for an $N=100$ spin ring and an initial
profile, schematically described by Fig.~\ref{fig11},
where $M=10$ spin vectors are specified as in Fig.~\ref{fig07}
and the remaining $90$ spin vectors are initially set anti-parallel
to the $3$-axis.
}
\end{figure}

\begin{figure}
 \includegraphics[width=\columnwidth]{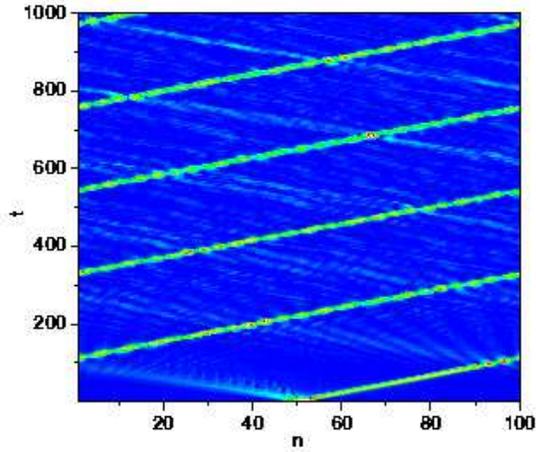}
\caption{\label{fig13}
Time evolution for an $N=100$ spin ring and an initial
profile, schematically described by Fig.~\ref{fig11},
where only $M=3$ spin vectors are specified as in Fig.~\ref{fig07}
and the remaining $97$ spin vectors are initially set parallel
to the $z$-axis. Even in this case a rather stable quasi-soliton
is created. However, with its generation a small localized perturbation
appears which moves in the opposite direction. Its interference with the
quasi-soliton can be observed several times (first at time $t\approx 100$).
}
\end{figure}

Based on the initial profile used for the soliton shown
in Fig.~\ref{fig08} we have used $M=10$ as well as $M=3$
for the corresponding truncated versions.
As shown in Figs.~\ref{fig12} and \ref{fig13}, soliton behavior is
clearly maintained for the choice $M=10$ and surprisingly,
despite the fact that this is a nonlinear problem,
even for $M=3$. For the case $M=3$ there is only an
increased level of background noise while the major
features of the quasi-soliton are retained.\\

\begin{figure}
 \includegraphics[width=\columnwidth]{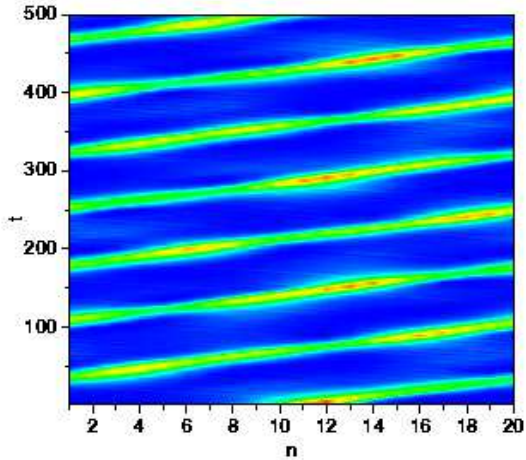}
\caption{\label{fig14}
Time evolution for an $N=20$ spin ring, where the initial data are selected
as described in the text. Soliton-like behavior persists although some
background noise and pulsation of the amplitude $z_n(t)$ is visible.
}
\end{figure}

\begin{figure}
 \includegraphics[width=\columnwidth]{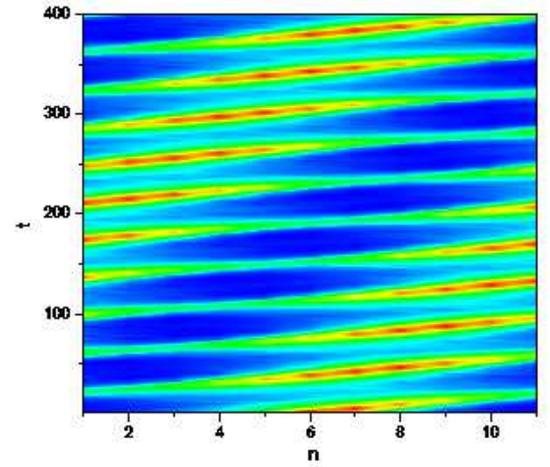}
\caption{\label{fig15}
Time evolution for an $N=11$ spin ring, where the initial data are selected
as described in the text. While some features of soliton-like behavior
are still visible, because of the small value of $N$ the results are strongly
affected by background noise and pulsation of the amplitude $z_n(t)$.
}
\end{figure}

\begin{figure}
 \includegraphics[width=\columnwidth]{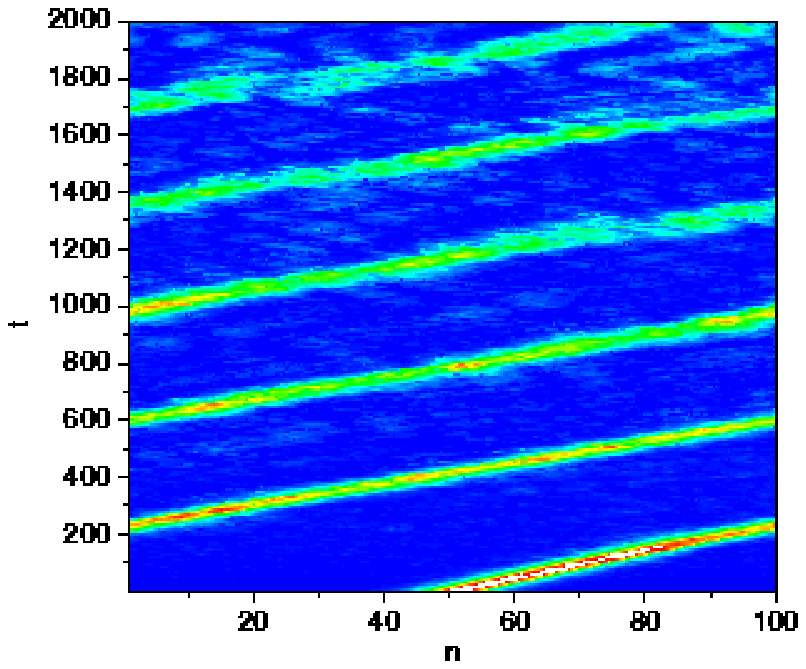}
\caption{\label{fig16}
Finite temperature effects on a quasi-soliton solution $z_n(t)$
for an $N=100$ spin ring and initial data calculated from
Eqs.~(\ref{E2}) - (\ref{E4}) using the parameters
$q=-8\pi/100$ and $u=\sin |q|$. Compared to the $T=0$ case
shown in Fig.~\ref{fig08} one observes that thermal fluctuations lead
to a decay of the soliton, however, it persists for a significant length
of time before eventually dissipating.
}
\end{figure}

While our quasi-soliton solutions appear to be robust
against {\it symmetric} truncation we have found that
{\it asymmetric} off-center truncation leads to far less
stable behavior, i.~e.~, the soliton decays after some time
which depends on the extent of the asymmetry.\\

Another way of truncating the initial profile is to
reduce the system size $N$, but leaving the initial profile intact.
This means, that for the system sizes where $N<M$ the truncation
effects are due to the imposed cyclic boundary conditions.
Our study is based on the numerical solution $z_n(t)$ with the
initial data calculated from Eqs.~(\ref{E2}) - (\ref{E4})
using the parameters $q=-32\pi/100$ and $u=\sin|q|$.
In Fig.~\ref{fig14} we show results for a spin ring with
$N=20$. The initial profile is symmetrically truncated
around the maximum value of $z_n(t)$. One clearly sees soliton
behavior with some background noise and a slight pulsation of the
amplitude $z_n(t)$. By reducing the system size to $N=11$ spins
soliton behavior is still visible, however the pulsation of the
amplitude is strongly enhanced (see Fig.~\ref{fig15}).\\

While in all of the above cases the total energy is a conserved quantity
we now show one specific example of the behavior of the quasi-solitons
when the spin system is coupled to a heat bath.
The behavior of classical spin systems in contact
with a heat bath can very effectively be studied
using a constant temperature stochastic spin dynamics approach,
such as that used in \cite{schroeder} \cite{ATH}.
Here the spin system is coupled to a heat bath
according to a Langevin-type approach by
including a Landau-Lifshitz damping term as
well as a fluctuating force with white noise characteristics.\\

Starting from prescribed initial conditions
as those used in Fig.~\ref{fig08}
the temperature is set to a value $T > 0$ and the trajectory
of the system is calculated numerically by solving a
stochastic Landau-Lifshitz equation, thereby
allowing the spin system to exchange energy
with its environment \cite{schroeder} \cite{ATH}.
Our results are shown in Fig.~\ref{fig16}.
Thermal fluctuations lead to a decay of the quasi-soliton,
driving the system towards its (ferromagnetic)
equilibrium configuration. However, the soliton persists
for a significant length of time before
eventually dissipating, as expected.\\

The cases considered in this Section provide a
picture of quasi-solitons in classical Heisenberg
rings that are remarkably robust even when
perturbed by truncations, small system size,
and heat bath couplings.

\section{Summary\label{sec:SU}}

The overriding theme of this paper is that relatively
small classical Heisenberg rings exhibit a rich variety
of dynamical properties including soliton behavior.
We have proposed that an effective strategy for
exploring the dynamical behavior of these rings
is to solve the system of discrete EOM of Eq.~(\ref{I1})
upon selecting initial spin configurations that are
discretized versions of the solutions of a new
continuum EOM that we derived by allowing for
continuous modulations of the spin wave solutions
given by Eq.~(\ref{I2}).  In practice, due to the
complexity of the analysis, we have limited our
analysis to a second order PDE version, Eq.~(\ref{SO1}),
of the continuous EOM of Eqs.~(\ref{D6}) and (\ref{D7}).
This second order PDE is a generalization of the
Landau-Lifshitz equation that has been used
extensively in studying one-dimensional magnetic
systems, e.~g.~, spin chains. Among the class of
solutions of Eq.~(\ref{SO1}) we have found soliton solutions.
Using discretized versions of the soliton solutions
as input data, we have found that the discrete
EOM of Eq.~(\ref{I1}) possess quasi-soliton solutions.
The robustness of these discrete quasi-solitons
has been demonstrated both by considering truncated
versions of continuum soliton solutions as well as by
investigating the dynamical behavior when the spins
are coupled to a heat bath at finite temperatures.
Apart from quasi-soliton behavior there is a vast
diversity of other dynamical behavior
exhibited by finite spin rings.
We have only sparsely investigated this diversity,
and at present we have only a limited physical
understanding of the behavior we have observed.
We conjecture that the existence of robust
quasi-solitons can be explained by the
existence of exact solitary solutions of
Eq.~(\ref{I1}), i.~e.~, localized solutions satisfying
$\mathbf{s}_{n+1}(t+T)=\mathbf{s}_n(t)$
for all $n=1,\ldots,\,N$ and some fixed $T$.\\

Magnetic solitons have been detected in many
magnetic systems through their signature on such
observables as specific heat,
NMR, and neutron scattering \cite{mikeska}.
Apart from any intrinsic interest in the dynamics of classical
Heisenberg spin rings, we were motivated to undertake the present
study as a first step in considering their quantum analogues.
This is especially relevant in view of the rapidly growing number
of quantum spin rings that are now available in the form
of synthesized magnetic molecules \cite{GSV} - \cite{W2}.
It is thus  timely to ask what is the quantum analogue of solitons
in ring systems, and how might these be detected by experiment \cite{SS}?\\

\section*{Acknowledgements}
Ch.~Schr\"oder is grateful to the
University of Applied Sciences Bielefeld  for
financial support.
Work at the Ames Laboratory was supported by the
Department of Energy-Basic Energy Sciences under
Contract No. DE-AC02-07CH11358. We thank
Paul Sacks of Iowa State University's
Department of Mathematics for useful discussions on
Lax pairs and ${\cal N}$-soliton solutions.
H.-J.~Schmidt thanks Ames Laboratory for funding an
extended visit to Ames where much of this research was performed.\\

\end{document}